\documentclass[12pt, letterpaper, oneside]{article}
\pdfoutput=1

\usepackage{graphicx}
\usepackage[body={6.5in, 9in}, right=1in, top=1in]{geometry}
\usepackage{amssymb}
\usepackage{amsmath} 
\usepackage{hyperref}
\usepackage[subnum]{cases}
\usepackage{color}

\newcommand\ea{\end{align}}
\newcommand\ba{\begin{align}}
\newcommand\ee{\end{equation}}
\newcommand\be{\begin{equation}}
\newcommand\eea{\end{eqnarray}}
\newcommand\bea{\begin{eqnarray}}

\newcommand\pd{\partial}
\newcommand\mpl{M_{\rm Pl}}

\newcommand\na{\nabla}
\newcommand\Tr{{\rm Tr}}
\newcommand\N{\mathcal{N}}
\newcommand\M{{\bf M}}

\begin{document}

\begin{center}
\LARGE{\textbf{Construction of the Conserved Non-linear $\zeta$ via the Effective Action for Perfect Fluids}} \\[1cm]
\large{Junpu Wang}
\\[0.4cm]

\vspace{.2cm}
\small{\textit{Department of Physics and ISCAP, \\ 
Columbia University, New York, NY 10027, USA}}

\end{center}

\vspace{.2cm}

\begin{abstract}

We consider the problem of how to construct the curvature perturbation $\zeta$ to nonlinear levels, which is expected to evolve time independently on super-horizon scales; in particular we concentrate on the situation where the universe is dominated by a perfect fluid.  We have used a low energy/long wavelength effective action to model the fluid sector.  Different from previous work, our approach assumes neither the absence of vector and tensor perturbations nor ``local homogeneity and isotropy''.  As a corollary, we also show that the nonlinearly defined graviton field $\gamma_{ij}$ is conserved outside the horizon in the same manner as $\zeta$ is.

\end{abstract}

%%%%%%%%%%%%%%%%%%%%%%%%%%%%%%%%%%%%%%%%%%%%%%%%%%%%%
%%%%%%%%%%%%%%%%%%%%%%%%%%%%%%%%%%%%%%%%%%%%%%%%%%%

\section{Introduction}
Despite the significant success of the linear perturbation theory in cosmology over the last few decades,  more and more efforts have been devoted in recent years to understanding the cosmological perturbations beyond the linear level --- both their origins and evolutions.  This is partially motivated by the extraordinary improvement in the accuracy of experiments; for instance, both the CMB data and the large scale structure data have indicated that there might be a detectable non-Gaussian feature in the primordial curvature perturbation.  Therefore to fully exploit the data, one has to study various cosmological perturbations at non-linear level.

Among these quantities, enormous attention has been paid to the curvature perturbation $\zeta$. Originally it was defined at linear order as the scalar metric perturbation on the uniform-density hyper-surfaces by Bardeen \cite{Bardeen} and was shown to evolve time-independently outside the horizon (cf. for instance \cite{Weinberg}). Many studies have been done recently in seek of extending the construction of a conserved $\zeta$ to non-linear level in different contexts.  Roughly speaking, there are three classes of methods employed to achieve that, which are summarized as follows: 
\begin{itemize}
\item The first one is the standard perturbative approach --- a perturbative expansion in fields \cite{MW}. It is straightforward and the equations governing the evolution of $\zeta$ are valid on all scales. However the price to pay is that the equations usually become cumbersome once one goes beyond the linear order in fields and there is a lack of systematic ways to demonstrate the conservation of $\zeta$ on super-horizon scales to arbitrary orders in the field expansion. 

\item The second class of approaches instead focuses on the spatial gradient expansion --- i.e. a expansion in powers of $\sigma\equiv k/a H$, where $k$ denotes the wavenumber, --- since the super-horizon evolution of $\zeta$ is usually of most interest. At the leading order, it corresponds to the separate universe approach.  \cite{Maldacena, CFKS} investigated the cases with an inflating background. There the authors showed that in single field inflation models, $\zeta$ can be defined as the (only) dynamical scalar metric perturbation in the unitary gauge (in which the matter field is demanded unperturbed) and that such a $\zeta$ was conserved to {\em all} orders in fields on super-horizon scales. The proofs were based on a Lagrangian formalism, for the matter contents in these cases were specified by explicit matter actions. 
%Along another direction,  the conservation of the non-perturbative $\zeta$ in long wavelength limit 
On the other hand, the cases with a general FRW background were studied in \cite{NS, LMS, RS} (known as the $\delta N$ formalism). It was shown that the conservation of the nonlinear $\zeta$ on large scales was assured if the pressure of the matter content was only a function of the energy density or if the matter content could be modeled by a single scalar field. As opposed to the aforementioned field theoretic formalism used in inflation scenarios, these analyses relied on the field equations of motion; in the context of the Einstein gravity the energy conservation alone would be sufficient.  

\item A third approach was proposed in \cite{LVprd, LVprl} by invoking the purely geometrical description of the curvature perturbation. The covariant quantity they constructed there would satisfy an exact, non-perturbative, valid-to-all-scale conservation equation and would eventually reduce to usual $\zeta$ on large scales.  The equivalence between this covariant formalism and the $\delta N$ formalism was established in \cite{SWY, Naruko}. 
\end{itemize}

In this paper, we will present a systematic way of constructing the nonlinear conserved $\zeta$ in a universe dominated by a perfect fluid.  A low energy effective action for an ordinary perfect fluid \cite{DGNR, ENRW} will be employed to model the matter content. So our method here is parallel to that used in the single field inflation cases \cite{Maldacena, CFKS}.  On the other hand, although this scenario was investigated to some extent in previous literature,  we believe that there is still some novelties in our approach: {\it i}) It does not {\em assume} that on a sufficiently large smoothing scale, the universe looks locally homogeneous and isotropic (like a FRW patch), as was a crucial assumption in \cite{WMLL, LMS}; in fact, such a feature, though can be {\em derived}, does not play a vital role in our construction. And {\it ii}) our approach works in the presence of vector and tensor perturbations. Moreover it enables us to construct the vector and tensor counterparts of $\zeta$, --- i.e. the nonlinear vector and tensor perturbation that evolve time-independently outside the horizon. \footnote{The conservation of the vector field does not contradict the usual intuition that vector modes decay in the absence of anisotropic stress tensor. As we will see later, due to the symmetry requirements of a fluid, any constant (in time) transverse vector field configuration is not physical and hence can be set to vanish, leaving only the decaying configuration.  }
%%%%%%%%%%%%%%%%%%%%%%%%%%%%%%%%%%%%%%%%%%%%%%%%%%
%%%%%%%%%%%%%%%%%%%%%%%%%%%%%%%%%%%%%%%%%%%%%%
\section{Setup}
We begin with a brief review of the low energy effective description of an ordinary perfect fluid system in Minkowskian spacetime,  which was proposed and developed in \cite{DGNR, ENRW,DHNS, ENPW}. For the purpose of this paper, we focus on a fluid with {\em no} conserved charges and neglect all the dissipative effects. The readers already familiar with our notations are welcome to skip directly to Section \ref{Fluid in cosmology}, in which the effective fluid action is used as the matter action in cosmological context and in which some formulas are derived in preparation for the construction of the conserved (nonlinear) quantities on large scales.

\subsection{Effective Theory for Perfect Fluids in Flat Spacetime}

Our goal is to construct a Lorentz invariant low energy effective theory for an ordinary perfect fluid system. For an ordinary fluid system with no conserved charges, we can specify as the long wavelength degrees of freedom the comoving coordinates of fluid volume elements, parametrized by $\phi^I$, with $I=1,2,3$. So at any fixed time $t$, the physical position occupied by each volume element is given by $\vec x(\phi^I, t)$. 
%For space filling fluid, there is a differeomorphism between the comoving coordinates $\phi^I$ and physical spatial %coordinates $x^i$ (with $i=,1,2,3$), for any fixed time $t$;  that is, we can use invertible functions 
%\footnote{In the sense that for any fixed $t$, the inverse functions $\phi^I(\vec x, t)$ exist. } 
%$x^i(\vec \phi, t)$ to denote the position of the fluid element, labeled by $\phi^I$, at instant $t$.
In this description, known as the Euclidean description, the physical spatial coordinates $x^i$ serve as dynamical fields while $t$ and $\phi^I$ are analogous to world sheet coordinates. 

However, it is often more convenient to use the inverse functions $\phi^I (\vec x,t )$ (\footnote{For space filling fluid, there is a differeomorphism between the comoving coordinates $\phi^I$ and physical spatial coordinates $x^i$ so that for any fixed time $t$, the inverse functions $\phi^I(\vec x, t)$ exist. }) as dynamical degrees of freedom (known as the Lagrangian description), since the spacetime symmetry can be straightforwardly implemented ---  we simply demand  $\phi^I$ to transform as scalars under Poincare transformations.  Also we are allowed to choose the comoving coordinates in such a manner that when the fluid system is at rest, in equilibrium and in a homogeneous state at some given external pressure, $\phi^I=x^I$ --- in the field theoretical language, this is equivalent to specifying the ground state of our theory to be
\be\label{ground state}
\langle \phi^I \rangle =x^I
\ee 

What are the other symmetries, in addition to the Poincare invariance, required to make the system behave like an ordinary fluid?  Notice that our choice of the ground state (\ref{ground state}) breaks both spatial translational and rotational invariance. In order that the energy momentum tensor of the system in equilibrium remains homogeneous and isotropic, as is indeed the case for a fluid, we shall impose {\it internal } symmetries to compensate the spontaneously broken {\it spacetime} symmetries. More precisely,  we demand that the theory be invariant under such internal transformations (i.e. all fields evaluated at the same spacetime point) as: 
\begin{align}
&\mathcal{T}_i:\;\phi^I\to \phi'^I=\phi^I +a^I,\quad a^I \text{constant} \label{Internal Translation}\\
&\mathcal{R}_i:\;\phi^I \to \phi'^I=O^I_{\;J}\phi^J,\quad O^I_{\;J}\in SO(3)
\end{align}
It is easy to show that the background configuration (\ref{ground state}) is invariant under the {\it diagonal} translation and rotation, respectively, which are defined as a linear combination of the spatial and internal transformations : $\mathcal{T}_d\equiv \mathcal{T}_s+\mathcal{T}_i$ and $\mathcal{R}_d\equiv \mathcal{R}_s+\mathcal{R}_i$; it is these residual symmetries that correspond to the homogeneity and isotropy of our background configuration.  

Moreover an ordinary fluid is insensitive to incompressional deformations --- it costs no energy to displace fluid volume elements if they are not compressed or dilated. Expressed in terms of a symmetry requirement, we demand that the theory be invariant under volume preserving differomorphisms of the comoving coordinates, 
%(understood as internal transformations)
defined as 
\be \label{Volume Preserving Diffeo}
\mathcal{D}: \;\phi^I\to \xi^I(\vec \phi ),\quad \text{with}\;\; \det \frac{\pd \xi^I }{\pd \phi^J}=1.
\ee

So now we are ready to construct the effective action for an ordinary fluid from $\phi^I$'s, compatible with the symmetry properties mentioned above. It is organized as a derivative expansion. Since the internal translation symmetry (\ref{Internal Translation}) mandates each field to be accompanied by at least one derivative, at the leading order in the derivative expansion,  the only invariant is
\be\label{def B}
B\equiv \det (B^{IJ}), \quad \text{with }\; B^{IJ}=\pd_\mu\phi^I\pd^\mu \phi^J
\ee
and hence the effective action takes the form of
\be
S_{\rm Fluid}=\int\! {\rm d}^4x \;F(B) \label{action}
\ee
where $F$ is a generic function, which, as we will see, characterizes the equation of state of the fluid in question. 

To illustrate that this effective action indeed describes a perfect fluid, we need to check that the (conserved) energy moment tensor takes the famous form: 
\be\label{EM tensor Fluid}
T_{\mu\nu}=(\rho+p)u_\mu u_\nu+p\,\eta_{\mu\nu}
\ee
Before doing that let's work out the four-velocity field $u^\mu(x)$ for the fluid system in our field theoretical language, which by definition is given by
\be
0=\frac{d}{d \tau} \phi^I (x)\equiv u^\mu(x)\pd_\mu \phi^I(x)
\ee
where $\tau$ parametrizes the streamline and the derivative with respect to $\tau$ vanishes because the comoving coordinate (or the label) of each fluid volume element is fixed. By solving the above equation for $u^{\mu}$, we obtain
\be
u^\mu=-\frac{1}{\sqrt{B}}\;\epsilon^{\mu\alpha \beta\gamma}\;\pd_\alpha \phi^1 \pd_\beta \phi^2\pd_\gamma  \phi^3 \label{u}
\ee
where $\epsilon$ is the 4d Levi-Civita symbol,
%\footnote{We save $\epsilon$ for later to denote Levi-Civita tensor: $\epsilon$}
with the convention $\epsilon_{\,0123}=-\epsilon^{\,0123}=1$. The normalization and overall sign of $u^\mu$ are chosen such that $u^\mu u_\mu=-1$ and $u^0>0$. 

The energy momentum tensor following from the effective action (\ref{action}) reads
\be
T_{\mu\nu}=-2F'(B)B (B^{-1})_{IJ}\;\pd_\mu\phi^I\pd_\nu\phi^J+\eta_{\mu\nu}F(B)\;.
\ee
With the aid of (\ref{u}), the energy momentum tensor above indeed can be recast into the perfect fluid form (\ref{EM tensor Fluid}), if we identify $\rho=-F(B)$ and $p=F(B)-2F'(B)B$.  This also justifies our preceding claim that the generic function $F$ determines the equation of state for fluids. For instance, for an ultra-relativistic fluid with $p=\rho/3$, one has $F(B)\propto B^{2/3}$. 

%We can also work out the equations of motion for the scalar fields:
%\be\label{eom flat metric}
%\pd_\mu\left(2F'(B)B(B^{-1})_{IJ}\pd^\mu \phi^J \right)=0, \quad I,J=1,2,3.
%\ee

%Not surprisingly, one can show that these equations provide {\it no} more information about the dynamics of the fluid system than the conservation of $T_{\mu\nu}$, which is consistent with our intuition from hydrodynamics. We will make frequent use of the former in the rest of this paper.  

In the rest of this subsection, we will consider small fluctuations about the homogeneous equilibrium background configuration (\ref{ground state}); they are associated with Goldstone excitations.
\be
\phi^I=x^I+\pi^I(x)\;. 
\ee 
Not all the $\pi^I$ fields feature propagating wave solutions. This can be seen by expanding the effective Lagrangian (\ref{action}) to the quadratic order: 
\be
\mathcal{L}^{(2)}=\frac{1}{2}w_0\left(\pi_L^2-c_s^2(\vec {\pd} \pi_L)^2\right)+\frac{1}{2}w_0\dot{\vec {\pi}}_T^2
\ee
where $\pi_L$ and $\vec{\pi}_T$ are the longitudinal (curl-free) and transverse (divergence-free) components of $\pi^I$: 
\be\label{pi decomposition}
\pi^I=\frac{\pd^I}{\sqrt{-\pd^2}}\pi_L+\pi_T^I
\ee
and where $w_0$ is the equilibrium enthalpy density ${\bar{\rho}+\bar{p}}$.  The (squared) speed of sound $c_s^2$ is given by
\be
c_s^2=\frac{d p}{d \rho}\Big \vert_{B=\bar{B}}=\frac{2F''(B)B+F'(B)}{F'(B)}\Big \vert_{B=\bar{B}}
\ee

Indeed, we see that only the longitudinal Goldstone field $\pi_L$ admits the standard propagating mode at a finite speed $c_s$, with a dispersion relation $\omega=c_s k$, while the other Goldstone field $\vec{\pi}_T$ has a degenerate dispersion relation $\omega=0$ and thus does not propagate. For this reason,  we usually interpret $\pi_L$ as the sound wave d.o.f. and $\vec{\pi}_T$ the vortex d.o.f. . 

It is worth pointing out that after spontaneous breaking, the spacetime symmetries get mixed with the internal ones, so the three Goldstone fields $\pi^I$ transform as a vector field under the diagonal $SO(3)$. \footnote{For this reason in the rest of the paper, we will not distinguish the spatial label ``$i,\,j,\dots$'' from the internal ones ``$I, \,J, \dots$''. } And after the decomposition (\ref{pi decomposition}), $\pi_L$ can be regarded as a scalar field and $\vec{\pi}_T$ as a transverse vector field.

%Before ending this subsection, it is worthy to remark the peculiar features of our effective action revealed when quantum effects are taken into account.  {\bf Bla...}
%%%%%%%%%%%%%%%%%%%%%%%%%%%%%%%%%%%%%%%%%%%%%%%%%%%%%%%%%%

\subsection{Cosmological Models with Perfect Fluids}\label{Fluid in cosmology}
In cosmology, the matter contents are often modeled as perfect fluids. Fortunately, in our field theoretic language, we can straightforwardly write the action of the cosmological model as 
\be \label{CosFluid}
S=S_{\rm EH}+S_{\rm Fluid}
\ee
where the first term on the r.h.s. is just the usual Einstein - Hilbert action for gravity and the matter action $S_{\rm Fluid}$ is given by the fluid effective action (\ref{action}), with the flat metric $\eta$ replaced by a cosmological spacetime one $g$ and the measure by $\sqrt{-g}\;{\rm d}^4 x$\;. We also assume that the metric fluctuates around a flat FRW background $\bar{g}_{\mu\nu}={\rm diag}\{-1,a(t)^2,a(t)^2,a(t)^2 \}$. The number of dynamical degrees of freedom in question is counted as follows. In addition to the three Goldstone fields $\pi^I$ --- which, as argued in last subsection, are eventually regrouped into one longitudinal scalar and one transverse vector under the residual $SO(3)$ group --- of the fluid sector,  the gravity sector introduces one extra $2-$polarized d.o.f. --- the spin$-2$ graviton, a traceless transverse tensor field. 

Perhaps a more transparent way to see this is by removing the gauge redundancy of the gravity sector. Using the ADM variables, one knows that only the spatial part of the metric is dynamical, which can be cast into the most general form of
\be\label{spatial metric}
g_{ij}\equiv h_{ij}=a(t)^2\exp{\left(A \delta_{ij}+ \pd_i \pd_j \chi +\pd_i C_j +\pd_j C_i +D_{ij}\right)}
\ee
with $C_i, D_{ij}$ satisfying $\pd_i C_i=0$ and $\pd_i D_{ij}=D_{ii}=0$. If the four gauge conditions of coordinate transformations are chosen to set $A=0,\; \chi=0, \;C_i=0$ (\footnote{We show in Appendix \ref{App SFSG} that this can always be done via a perturbative construction. }) --- which is known as the spatially flat slicing gauge (SFSG), --- we are left with $\pi_L,\; \vec{\pi}_T,\;$ and  $D_{ij}$ as the dynamical degrees of freedom, which characterizes, respectively, the scalar,  vector and tensor cosmological perturbation.  

For the interest of this paper, it turns out to be most convenient to work in another gauge --- called the unitary gauge (UG), --- in which all the perturbations are absorbed into the metric, leaving the matter fields unperturbed: $\phi^I=x^I$. Said differently, the spatial coordinates are chosen to coincide with the comoving ones. Meanwhile the temporal gauge freedom is used to determine the time slices such that the scalar quantity $B$ (defined in eqn. (\ref{def B})) remains unperturbed on each time slice: $B(t)=a(t)^{-6}$. 

The unitary gauge leads to many conceptual and computational simplifications.  First of all,  with our choice of the spatial coordinates, the worldlines of fluid volume elements coincide with the threads $x^i=$ constant. Indeed, this can be seen by considering the spatial components of the velocity field, which vanish since $u^I \propto \epsilon^{I a b c}\,\pd_a \phi^1\,\pd_b \phi^2\,\pd_c\phi^3=0$.  Second, our choice of time slices coincides with the uniform density slices, for the energy density of the fluid, given by $\rho=-F(B)$, is only a function of $B$ and it is a constant on each time slice.  Using the ADM variables, we can parametrize the metric as
\be
ds^2=-\mathcal{N}^2 dt^2+h_{ij}(d x^i+N^i dt)(d x^j+N^j dt)
\ee 
and the inverse metric $g^{\mu\nu}$ as
\be
g^{00}=-\frac{1}{\mathcal{N}^2},\quad g^{0i}=g^{i0}=\frac{N^i}{\mathcal{N}^2},\quad g^{ij}=h^{ij}-\frac{N^i N^j}{\mathcal{N}^2}
\ee
where the spatial metric $h_{ij}$ takes the form of (\ref{spatial metric}) and $h^{ij}$ is the inverse {\it spatial} metric: $h^{ik}h_{kj}=\delta ^i _j$. One can show easily that in the UG $B^{IJ}=g^{IJ}$ and that
\be
B=\det g^{IJ}=\frac{1}{\det h_{IJ}}\left(1-\frac{N^I N_I}{\mathcal{N}^2}\right),
\ee  
where the index of $N^I$ ($N_I$) are lowered (raised) by $h_{IJ}$ ($h^{IJ}$). 
%Thus using (\ref{spatial metric}) and identities in matrix analysis
%\begin{align*}
%&\det\left(\delta^i_{\;j}-v^i v_j\right)=1-v^i v_i\quad \text{and }
%\det {\rm e}^{M^i_{\;j}}={\rm e}^{M^i_{\;i}}
%\end{align*}
%Not surprisingly the velocity field of fluid (in curved spacetime) is reduced to 
%\begin{subnumcases}
%{u^a\equiv -\frac{1}{\sqrt{B}}\epsilon^{abcd}\na_b \phi^1\na_c \phi^2\na_d \phi^3=}
%\frac{1}{\sqrt{N^2-N^IN_I}}& for $ a=0 $\\
%0 & for $a=1,2,3$
%\end{subnumcases}
%where $\epsilon^{abcd}$ is the Levi-Civita tensor defined as $\epsilon^{abcd}=\tilde{\epsilon}^{abcd}/\sqrt{-g}$. 
Therefore the time slicing condition can be expressed as
\be \label{UG time slicing}
B(t)=a(t)^{-6}\;\iff\;3A+\nabla^2 \chi=\log\left(1-\frac{N^I N_I }{\mathcal{N}^2}\right).
\ee

Now the number of dynamical d.o.f.  in the UG can be counted as follows: since $\mathcal{N}$ and $N^I$ are just auxiliary fields and can be determined in terms of $h_{ij}$ after algebraically solving the imposed the constraints, (\ref{UG time slicing}) implies that the two scalar functions $A$ and $\chi$ are not independent. So henceforth we shall always regard $\chi$ as being expressed (perturbatively) in terms of other perturbations in the metric and select $A$, $C_i$ and $D_{ij}$ as the dynamical fields, the number of which is in agreement with that in the SFSG. 

We are ready to solve in the UG the constraint equations $\delta S/\delta N=0, \delta S/\delta N^i=0$ and the time slicing condition (\ref{UG time slicing}) up to the linear order in fields and to obtain the UG quadratic action: 
\begin{align}
S_T^{(2)}&=\int dt \int_{\vec k}\,\frac{\mpl^2}{8}\,a(t)^3\,\left(\dot{D}_{ij}^2-\frac{k^2}{a^2}D_{ij}^2\right),\label{Fluid Qua Tensor}\\
S_V^{(2)}&=\int dt \int_{\vec k}\,a(t)^3\,\frac{-\mpl^2\dot{H}a(t)^2}{1-4\dot{H}a^2/k^2}\,\dot{C}_{i}^2,\label{Fluid Qua Vector}\\
S_S^{(2)}&=\int dt \int_{\vec k}\,a(t)^3\,\left[\frac{9\mpl^2\dot{H}a(t)^2}{4(3\dot{H}a^2-k^2)}\,\left(\dot{A}-\frac{\dot{H}}{H}\,A\right)^2-\frac{9\mpl^2\dot{H}}{4}\left(1+\frac{\ddot{H}}{3H\dot{H}}\right)\,A^2\right].\label{Fluid Qua Scalar}
\end{align}
where the symbol $\int_{\vec k}$ is abbreviation for $\int \frac{d^3 k}{(2\pi)^3}\,$ and where we have used the Friedmann equation 
\be
\bar{\rho}=-F(\bar{B})=3\mpl^2H^2,\quad \bar{\rho}+\bar{p}=-2F'(\bar{B})\bar{B}=-2\mpl^2 \dot{H}.
\ee
In the long wavelength limit $\sigma \equiv k/aH \to 0$, the actions (\ref{Fluid Qua Tensor}---\ref{Fluid Qua Scalar})reduce to
\be \label{IR qua action}
S^{(2)}=S_T^{(2)}+S_V^{(2)}+S_S^{(2)}\simeq \int dt \int_{\vec k}\,\mpl^2\,a(t)^3\,\left[\frac{1}{8}\,\dot{D}_{ij}^2+\frac{1}{4}\left(k\dot{C}_i\right)^2+\frac{3}{4}\,\dot{A}^2\right]
\ee

A few comments are in order: 
{\it i}) The (squared) speeds of sound for fluids in curved spacetime are $c_s^2=d p/d\rho\vert_{\bar{B}}=-1-\ddot{H}/3H\dot{H}$ and $c_T^2=0$, respectively, for the longitudinal and transverse Goldstone excitations.  For $c_s$ to be sub-luminal, we need $1<-\ddot{H}/3H\dot{H}<2$; for fluids with constant $w=p/\rho$, this condition corresponds to the scale factor evolving as $a(t)\propto t^n$, with $\frac{1}{3}<n<\frac{2}{3}$. In particular, this implies that our cosmological model (\ref{CosFluid}) with a perfect fluid can not be a {\em consistent} inflationary model. Indeed to have inflation, we need to relax some symmetry requirements for our system, using less symmetric object (e.g. a solid \cite{ENW}) as the matter content.  {\it ii}) The IR quadratic action (\ref{IR qua action}) implies that we should really treat $A\sim k C_i\sim D_{ij}$ as of the same order in the spatial gradient expansion. Also despite the deceptive appearance, $\na^2 \chi$ is not necessarily of higher order than $A$ in the spatial gradient expansion; in fact as we will show soon, they are both of the leading order. Thus the scalar, vector and tensor metric perturbation in (\ref{spatial metric}) are all of the same order in the spatial gradient expansion, i.e.
\be
A\sim k^2\chi\sim k C_i\sim D_{ij} \sim \mathcal{O}\left(\sigma^0\right)
\ee
%Finally, we show that the e.o.m.~for the Goldstone fields (\ref{eom flat metric}) in the UG get significantly simplified as well.  

%%%%%%%%%%%%%%%%%%%%%%%%%%%%%%%%%%%%%%%%%%%%%%%%%%%%%%%%
%%%%%%%%%%%%%%%%%%%%%%%%%%%%%%%%%%%%%%%%%%%%%%%%%%%%%%%
\section{Construct the Conserved Curvature Perturbation to Non-linear Order}
As mentioned in the previous section,  if the matter content of the universe can be modeled as an ordinary perfect fluid, there is only one dynamical scalar field. In the UG it is parametrized by $A(x)$, the (dynamical) scalar perturbation in the spatial metric (eqn. (\ref{spatial metric})). In this section, we will show that the evolution of $A$ will remain time-independent as long as the mode is outside the horizon.  The conservation is preserved up to {\it all} orders in the field expansion; in particular, at the linear order $A$ coincides with (twice) the usual (linear) curvature perturbation. Thus we shall define our linear curvature perturbation $\zeta$ by $\zeta=A/2$. 

%{\bf Summarize the steps of proof. }
The proof of the conservation of $\zeta$ on large scales will proceed in several steps as follows: Firstly we expand the action (\ref{CosFluid}) to the leading order in the spatial gradient, i.e. to the order $\mathcal{O}(\sigma^0)$, while keeping all orders in fields. We show that the IR action starts with terms involving at least two time derivatives,  each of which acts on a different field, and hence the constant (time-independent) configurations of $A, C_i, D_{ij}$ are permitted as solutions to the classical equation of motion.  And secondly we show that these constant configurations are stable under small fluctuations --- i.e. they are indeed attractors. As we will see the proof for the scalar, vector and tensor perturbations are essentially identical.

%\vspace{5pt}
\noindent {\it Step 1}: To show the conservation of $A,C_i, D_{ij}$, we expand the Lagrangian (\ref{CosFluid}) up to the {\it first} order in the temporal derivative and to the {\it zeroth} order in the spatial gradients, while keeping all orders in fields.  The main challenge of doing this is to express the non-dynamical quantities, such as $\N,N^i$ and $\chi$, in terms of the dynamical ones, by using the time slicing condition (\ref{UG time slicing}) and the constraint equations which are given by
\begin{align}
0=\frac{\delta S}{\delta \N}=&\frac{\mpl^2}{2}\Bigg[R^{(3)}+\N^{-2}\Bigg(6H^2+2H \,\Tr\,\dot{\M}+\frac{1}{4}\Tr\left(\frac{d}{dt }e^{-\M}\frac{d}{dt }e^{\M}\right)
+\frac{1}{4}\left(\Tr \dot{\M}\right)^2\nonumber\\&
+\hat{\na}_i N^j (\dots)\Bigg)\Bigg] 
+F(B)+2F'(B)B\frac{N^i N_i}{\N^2-N^i N_i}\label{constraint1}\\
0=\frac{\delta S}{\delta N^i}=&\hat{\na}_i \left[-\frac{2H}{\N}-\frac{\Tr\,\dot{\M}}{2\N}+\frac{\hat{\na}_k N^k}{\N}\right]+\hat{\na}_j\Bigg[\frac{1}{2\N}\left(e^{-\M}\frac{d}{dt}e^\M\right)^j_{\;i}-\frac{\hat{\na}^jN_i+\hat{\na}_iN^j}{2\N}\Bigg]\nonumber\\
&-2F'(B)B\frac{\N N_i}{\N^2-N_k N^k}\label{constraint2}
\end{align}
where $R^{(3)}$ is the {\it spatial} Ricci scalar constructed from $h_{ij}$, $\hat{\na}$ the covariant derivative compatible with $h_{ij}$, and the matrix $\M_{ij}=\log(h_{ij}a^{-2})$. The ``$\dots$'' in the first equation stands for terms that are regular in the limit $\sigma\to 0$ --- the form of which is irrelevant in the analysis. Inspecting the constraint equation (\ref{constraint2}), one finds that $N^i$ starts {\em at least} at the order $\mathcal{O}(\sigma)$. Thus, as long as the super-horizon modes are concerned, all the terms in (\ref{constraint1}) with $N^i$ or $\hat{\na}$ (as well as $R^{(3)}$) can be neglected, which leads to a crucial fact that $\delta \N\equiv \N -1 $  in the super-horizon regime starts with terms involving two time derivatives:
\be \label{delta N}
\delta \N=\left[1+\frac{1}{24H^2}\Tr \left(\frac{d}{dt}e^{-\M} \frac{d}{dt}e^{\M}\right)\right]^{1/2}-1,\qquad \sigma\to 0
\ee
Moreover, we see that for each field, there is at most one time derivative acting on it. 

On the other hand, the time slicing condition (\ref{UG time slicing}), in this long wavelength limit, reduces to 
\be
\Tr \, \M =3A+\na^2 \chi =\mathcal{O}\left(\sigma^2\right),
\ee
which verifies our claim that $\na^2 \chi$ is of the same order as $A$ in the spatial gradient expansion. 

It then follows immediately that in the super-horizon regime, the Lagrangian (\ref{CosFluid}) becomes
\be
\lim_{k/aH\to0}\mathcal{L}\simeq 2a(t)^3\N F(B)\simeq -6a(t)^3\mpl^2H^2+\sum_{n\ge 2} \mathcal{L}_n (\phi_a)
\ee
In the last step we have denoted the dynamical fields --- $A, C_i, D_{ij}$--- collectively by $\phi_a$\ and the number of fields contained in $\mathcal{L}_n$ by the subscript ``$n$''. Notice that schematically $\mathcal{L}_n$'s take the form of 
\be
\mathcal{L}_2\sim Q_2(t) G_2^{ab}\dot{\phi_a}\,\dot{\phi_b},\quad \mathcal{L}_{n>2}\sim Q_n(t)G_n^{ab\dots kl}(\phi)\dot{\phi_a}\,\dot{\phi_b}\dots \dot{\phi_k}\,\dot{\phi_l} 
\ee
with $Q_n(t)$ being a function of time consisting of $a(t), H$ etc.. Besides the irrelevant field-independent term,  this long-wavelength Lagrangian starts at {\em two} time derivative level and hence the e.o.m. following from it reads 
\be
f_1(\phi,\dot{\phi})\ddot{\phi_a}+f_2(\phi,\dot{\phi})^b_{\,a}\dot{\phi_b}=0
\ee
%with each field acted on by at most one time derivative (which can be seen via eqn (\ref{delta N})). 

Therefore it indeed admits $A,\,C_i,\,D_{ij}=$ constant as solutions to the classical equation of motion. Since this IR Lagrangian contains all orders in fields, the conservation of these fields on large scales must be preserved nonlinearly. 

\noindent {\it Step 2}: Now we show the solutions $A, \,C_i,\,D_{ij}=$ constant are actually attractors.  We just work on the scalar case, since the analysis for the other two is identical. Writing $A$ as $A=A_0+\delta A$, owing to the constancy of $A_0$, the quadratic action for the fluctuation $\delta A$ in the long-wavelength limit takes the same form as that for $A$, which is given by 
\begin{equation*}
\delta S_s^{(2)}=\int dt \int_{\vec k}\,\frac{3\mpl^2}{4}\,a(t)^3\,\dot{\delta A}^2\\
\end{equation*}
%where the subscript ``s'' is the abbreviation for ``scalar field'' and the symbol $\int_{\vec k}$ is for $\int \frac{d^3 k}{(2\pi)^3}\,$.  
from which follows the linearized equation of motion for $\delta A$
\be
\ddot{\delta A}+3H\,\dot{\delta A}=0
\ee
It admits two general solutions --- one decaying mode and one constant mode: 
% {\bf what if $a(t)\propto t^n$ with $n<1/3$} 
\be
\delta A_1=\int \frac{dt }{a(t)^3},\quad  \delta A_2=\text{const. }
\ee 
This thus confirms that the solution $A=$ constant is an attractor. 

%As a bonus, by applying the analysis in this section to the vector perturbation $C_i$ and the tensor perturbation $D_{ij}$,  we can show that they are also conserved in the super-horizon regime in the same manner as their scalar counterpart $A$. 
As we saw, our approach is more powerful in some aspect than those in previous literature, for it enables us to show that the vector perturbation $C_i$ and the tensor perturbation $D_{ij}$ are also conserved on super-horizon scales in the same manner as their scalar counterpart $A$. And in next section, we will show that the nonlinear, conserved $\zeta\equiv A/2$ and $D_{ij}$ we constructed here agrees with that in \cite{LMS,MW}, up to a global spatial coordinates redefinition. 

%%%%%%%%%%%%%%%%%%%%%%%%%%%%%%%%%%%%%%%%%%%%%%%%%%%%%%%
%%%%%%%%%%%%%%%%%%%%%%%%%%%%%%%%%%%%%%%%%%%%%%%%%%%%%%%
\section{Discussion}
We have constructed the curvature perturbation $\zeta$ (to nonlinear level) via the following steps: {\it i}) choose a coordinate system such that the spatial coordinates comove with the fluid and that equal time slice coincides with the uniform density slice.  And {\it ii}) define $\zeta$ to be (half) the coefficient of the term in $\log\left(a^{-2}g_{ij}\right)$ that is proportional to $\delta_{ij}$. That is,
\be
\zeta=\frac{1}{4}\na^{-2}\left(\delta_{ij}\na^2-\pd_i\pd_j\right)\log \left[a^{-2}g_{ij}\right]
\ee
We showed that the $\zeta$ such defined evolves time-independently in the super-horizon regime to all orders in fields. 

Our result of the conservation of the non-linear curvature perturbation $\zeta$ on large scales agrees with previous literatures \cite{LMS, MW}. But in our analysis we did not neglect the presence of the vector and tensor perturbations, nor did we use the property of local homogeneity and isotropy.\footnote{In fact having fixed the UG by $\phi^I=x^I$ and $B=a(t)^{-6}$, we have depleted all guage freedoms for choosing a coordinate system. Furthermore, as we argued earlier, we should not think the terms in $g_{ij}$ with ``superficially'' more derivatives such as $\pd_i\pd_j\chi, \;\pd_i C_j$ are of higher order in the spatial gradient expansion than $A\,\delta_{ij}$. As a consequence of which,  as long as the UG is chosen, we are not entitled to claim any property about $g_{ij}$ on large scales. }
As we showed explicitly, the spatial metric in UG on large scales ($\sigma\to 0$) is given by 
\be
g_{ij}= a(t)^2 \exp\left[6\zeta\left(\frac{1}{3}\delta_{ij}-\hat{k}_i\hat{k}_j\right)+\text{vector}+\text{tensor}\right].
\ee
%One then might think that even though the vector and tensor perturbations are neglected, this metric is anisotropic.  However, this puzzle stems from a non-conventional choice of coordinates: at the linear order in fields, there is a further coordinate transformation (from the UG) $x^I\to x^I -\frac{3}{2}\frac{\pd^I}{\na^2}A(\vec x)$, such that the term in $g_{ij}$ that is proportional to $\hat{k}_i\hat{k}_j$ gets eliminated, while keeping the spatial threading and time slicing unaltered. This is consistent with our intuition (about the linear perturbation theory of cosmological fluids) that no anisotropy in metric will be generated if the fluid is free of anisotropy stress.  However, beyond the linear level, it becomes much more involved: it is not clear to us whether there exists still such coordinate transformation to eliminate anisotropy in metric (terms proportional to $\hat{k}_i\hat{k}_j$) to all orders in fields, that works to all orders in fields. 
%Thus whether the spatial metric has a vanishing $\hat{k}_i\hat{k}_j$ piece nonlinearly is still an open question. 
At a glimpse, one may think that even though the vector and tensor perturbations are neglected, this metric is anisotropic, which is inconsistent with our intuition about cosmological fluids that no anisotropy in the metric will be generated if the fluid is free of anisotropic stress. However, this puzzle results from our non-conventional choice of coordinates: we can perform a further coordinate transformation (from the UG) 
\be\label{coo trans}
t\to t, \quad \;x^I\to y^I= x^I+\xi^I(\vec x)
\ee
such that 
\be \label{new coo system}
\phi'^I (t,\vec{y})=y^I+\xi'^I(\vec y),\quad \text{and } \;g_{ij}(t,\vec y)=a(t)^2 e^{2\zeta'(\vec y)}\exp \gamma'_{ij}(\vec y), \quad \text{with }\; \pd_i\gamma'_{ij}=\gamma'_{ii}=0
\ee
where the time-independence of $\xi^I (\vec x)$ follows directly from the time-independence of $A,C_i,$ and $D_{ij}$ and where the function $\xi'^I(\vec y)$ is obtained by inverting the 3-diffeomorphism $y^I(\vec x)=x^I+ \xi^I(\vec x)$. Notice that the coordinate transformation like (\ref{coo trans}) alters neither the time slices nor the threading of the spatial coordinate, the latter of which is because
\be
u'^I\propto \epsilon^{I\alpha \beta\gamma}\pd_{\alpha}\phi'^1\pd_{\beta}\phi'^2\pd_{\gamma}\phi'^3=0, \quad u'_I=g_{0I}u^0=h_{ij}N^j u^0\sim \mathcal{O}(\sigma).
\ee
Therefore this implies that in the new coordinate system and in the super-horizon regime, both the spatial metric $g_{ij}$ and $T_{ij}$ (c.f. eqn. (\ref{EM tensor Fluid}))--- which is reduced to $T_{ij}=p g_{ij}$ on large scales --- are isotropic, were the tensor perturbation ignored, and homogeneous since in the limit $\sigma\to0$ the spatial dependent of $\zeta'$ and $\gamma'$ is negligible. 

That is, although convenient for computational purposes, the UG coordinate is not capable of exhibiting the local homogeneity and isotropy in the long wavelength limit. Fortunately there exists a new coordinate system, which is indistinguishable from the UG coordinate by only inspecting the physical quantities such as $\rho, p, u^i$ etc.~and in which the existence of local homogeneous and isotropic patches on large scales can be verified (rather than being assumed, as were in previous literatures).  The $\zeta'$ in (\ref{new coo system}) is also time-independent on large scales and it agrees perfectly with the definition in \cite{LMS, MW}.

As we showed in last section, on super-horizon scales, the vector field $C_i$ evolves time-independently.  However, this does {\it not} correspond to a physical vector mode, as can be understood as follows. 
%Given the time independence of the metric perturbations $A, C_i, D_{ij}$ in UG on large scales,  we can perform a change of coordinate system 
%\be
%t\to t,\quad x^I\to y^I(\vec x)
%\ee
%to transform from the UG to the spatially flat slicing gauge,
%\be
%\text{UG: } \left\{ 
%\begin{array}{rl} \phi^I &=x^I\\g_{ij}&=a^2\exp\left\{(3\delta_{ij}-\frac{\pd_i\pd_j}{\na^2})A(\vec x)\right\} 
%\end{array}\right.
%\ee
%in which the configurations of the fluid are given by
%\begin{equation}\label{SFSG Fluid Config}
%\phi'^I (t,\vec{y})=x^I(\vec y)
%\end{equation}
%where the $x^I(\vec y)$ is the inverse function of $y^I(\vec x)$.   We don't need the explicit form of this %transformation except to notice that $\det \pd x^I/\pd y^J =1$. This is because
%\be
%g^{\text{SFSG}}_{ij}(t,\vec{y})=\frac{\pd x^\mu}{\pd y^i}\frac{\pd x^\nu}{\pd y^j}g^{\text{UG}}_{\mu\nu}(t,%\vec{x})=\frac{\pd x^k}{\pd y^i}\frac{\pd x^l}{\pd y^j}g^{\text{UG}}_{kl}(t,\vec{x})
%\ee
%and because $\det g^{\text{SFSG}}_{ij}\simeq\det g^{\text{UG}}_{ij}=a(t)^6$ on large scales.  
Transform to the coordinate system specified by (\ref{new coo system}). Owing to the internal volume-preserving-diffeomorphism invariance (\ref{Volume Preserving Diffeo}) of the fluid system, the configuration in eqn (\ref{new coo system}) is in fact symmetric to the one free of vortex degrees of freedom (divergence-free vector modes), either in the matter fields or in the dynamical metric components. That is, as the leading contribution, the constant (in time) $C_i$ configuration has no physical significance; the well known decaying vector modes in the universe dominated by perfect fluids come from subdominant, time-dependent terms in $g_{ij}$ in the UG. 
%hence there is no vector mode left in matter sector or in gravity sector. 

It is known that a perfect fluid free of vortex degrees of freedom (divergence-free vector modes) has an equivalent effective description involving only {\it one} scalar field  --- i.e.~the $P(X) \equiv P((\pd\psi)^2)$ description \cite{DGNR,BCNV, NSW}. Then we can apply the method in Ref.~\cite{Maldacena, CFKS} directly to a non-inflating background and define the curvature perturbation $\zeta$ as the (only) dynamical scalar metric perturbation in unitary gauge: 
\be
\psi(t,\vec x)=\psi_0(t),\quad g_{ij}=a^2 e^{2\zeta}\exp\{\gamma_{ij}\}
\ee
Details of the proof of the conservation of $\zeta$ on super-horizon scales in the $P((\pd\psi)^2)$ context are collected in Appendix \ref{P(X) Fluid}. The nonlinear $\zeta$ constructed via this logic is identical to the $\zeta'$ field in (\ref{new coo system}), hence agrees with our definition up to a global reshuffle of the spatial coordinates.  

\section{Conclusions}
In this paper we consider the problem of how to extend the definition of curvature perturbation $\zeta$ to nonlinear levels; in particular we concentrate on the situation where the universe is dominated by a perfect fluid.  We have used a low energy/long wavelength effective action to model the fluid sector and constructed the nonlinear $\zeta$ as follows:  {\it i)} Fix the gauge such that the comoving coordinates of the fluid coincide with the spatial ones and that the constant time slices are the uniform density slices; and {\it ii)} define $\zeta=A/2$ where $A$ is the isotropic scalar part in the spatial metric (\ref{spatial metric}).  We have shown this $\zeta$ (as well as its tensor counterpart) is conserved outside the horizon to all orders in fields. 

Although this topic has been investigated to some extent in previous literature, we believe our approach is novel and has its own merits. For instance, in discussing the conservation on super-horizon scales of $\zeta$ (the scalar mode), we do not need to neglect the vector or tensor modes;  on the contrary, we illustrate that there are vector (non-physical) and tensor counterparts of $\zeta$ which are conserved in the same manner as $\zeta$ is.  Moreover, our proof does not assume/rely on that the universe looks locally like FRW patches (local homogeneity and isotropy) on sufficiently large scales, which was the key input in previous work.  Nevertheless, we have shown our definition of $\zeta$ agrees with that in Lyth et al.~(2005) and Malik et al.~(2004) up to a global reshuffle of the spatial coordinates .

\section*{Acknowledgments}

We would like to thank Lam Hui and Alberto Nicolis for illuminating discussions and guidance. 
%We would also like to thank Solomon Endlich for useful discussions and for his careful reading of the manuscript.
 The work of J.~W.~is supported by the DOE under contract DE-FG02-92-ER40699.

\appendix
\section*{Appendix}
\section{the Spatially Flat Slicing Gauge (SFSG) to All Orders in Fields}\label{App SFSG}
In this section, we show that starting from an arbitrary $g_{ij}(x)=\bar{g}_{ij}+\delta g_{ij}$, we can always achieve the non-linear SFSG, specified by 
\be\label{Nonlinear SFSG}
g_{ij}(x)=a(t)^2 \exp \gamma_{ij}(x), \quad \text{with } \gamma_{ii}=\pd_i\gamma_{ij}=0
\ee
via some appropriate gauge transformation $x^\mu \to y^\mu=x^\mu+\xi^\mu(x)$.  Here we only concentrate on the most relevant case to our analysis, in which $\bar{g}_{\mu\nu}={\rm diag} \{-1,a(t)^2,a(t)^2,a(t)^2\}$ and $\delta g$ is treated as small fluctuations around the FRW background.  We prove this by using a perturbative construction. 

%First notice that $g_{ij}(x)$ can always be decomposed as
%\be\label{general spatial metric}
%g_{ij}=a(t)^2\exp\{A\delta_{ij}+\pd_i\pd_j\chi+\pd_i C_j+\pd_j C_i+ D_{ij}\}
%\ee
%with $C_i$ and $D_{ij}$ satisfying $\pd_i C_i=D_{ii}=\pd_i D_{ij}=0$. (\footnote{To see this, we define $A$ and $\chi$ to be the solutions of 
%\be
%3A+\na^2\chi=\Tr \log \left(a^{-2}g_{ij}\right) \quad \text{and} \quad \na^2 A+\na^2 \na^2 \chi=\pd_i\pd_j %\log(a^{-2}g_{ij}), 
%\ee
%then define $C_i$ as the solution of 
%\be
%\pd_j(A+\na^2\chi)+\na^2 C_j=\pd_i \log (a^{-2}g_{ij})
%\ee
%and then use eqn. (\ref{general spatial metric}) to determine $D_{ij}$.
%})
First let us invert the function $y^\mu(x)$ and denote it as 
\be
x^\mu(y)=y^\mu+v^\mu(y), \quad \text{with} \;v^\mu(y)=\sum_{n=1}^\infty v^\mu_{(n)}(y)
\ee
where $v^\mu_{(n)}$ is assumed to be of order $(\delta g)^n$. Under this gauge transformation, the spatial metric transforms as
\begin{align}
g_{ij}(x)\to \tilde{g}_{ij}(y)=\frac{\pd (y^\alpha+v^\alpha)}{\pd y^i}\frac{\pd (y^\beta+v^\beta)}{\pd y^j}g_{\alpha \beta}(y+v)
\end{align}
Consider $M_{ij}=\log(a^{-2}g_{ij})$. At the first order in $\delta g$, it is straightforward to work out $\tilde{M}^{(1)}_{ij}(y)$, which is given by 
\be\label{first order M}
\tilde{M}^{(1)}_{ij}=a^{-2}\delta g_{ij}+\pd_i v^j_{(1)}+\pd_j v^i_{(1)}+2H v^0_{(1)}\delta_{ij}
\ee
Notice that any tensor function $\delta g_{ij}(x)$ ($i,j=1,2,3$) can always be put in the form of 
\be\label{general tensor function}
\delta g_{ij}=a(t)^2 \left(\mathcal{A}\delta_{ij}+\pd_i\pd_j\mathcal{B}+\pd_i \mathcal{C}_j+\pd_j \mathcal{C}_i+ \mathcal{D}_{ij}\right)
\ee
with $\mathcal{C}_i$ and $\mathcal{D}_{ij}$ satisfying $\pd_i \mathcal{C}_i=\mathcal{D}_{ii}=\pd_i \mathcal{D}_{ij}=0$. \footnote{To see this, we define $\mathcal{A}$ and $\mathcal{B}$ to be the solutions of 
\be
3\mathcal{A}+\na^2\mathcal{B}=a^{-2}\delta g_{ii} \quad \text{and} \quad \na^2 \mathcal{A}+\na^2 \na^2 \mathcal{B}=a^{-2}\pd_i\pd_j \delta g_{ij}, 
\ee
then define $\mathcal{C}_i$ as the solution of 
\be
\pd_j(\mathcal{A}+\na^2\mathcal{B})+\na^2 \mathcal{C}_j=a^{-2}\pd_i \delta g_{ij}
\ee
and then use eqn. (\ref{general tensor function}) to determine $\mathcal{D}_{ij}$.
}
Thus by choosing 
\be
v^0_{(1)}=-\mathcal{A}/2H, \quad v^S_{(1)}=-\mathcal{B}/2,\quad v^i_{T(1)}=-\mathcal{C}_i
\ee
where $v^S$ and $v^i_T$ are the longitudinal (curl-free) and transverse (divergence-free) components of $v_i$: $v^i_{(n)}=\pd_i v^S_{(n)}+v^i_{T(n)}$, we find that $\tilde{M}^{(1)}_{ij}=\mathcal{D}_{ij}$ --- i.e. only the transverse traceless part survives. 

At the second order the expression for $\tilde{M}^{(2)}_{ij}$ is quite cumbersome, but that doesn't bother us much. Since it takes the form of 
\be
\tilde{M}^{(2)}_{ij}=\mathcal{S}_{ij}+\pd_i v^j_{(2)}+\pd_j v^i_{(2)}+2H v^0_{(2)}\delta_{ij},
\ee
by the same argument as above, we show that all the fields in $\mathcal{S}_{ij}$ can be set to zero except the TT one.  And this process can be repeated to all orders in $\delta g$.  

Now let $\gamma_{ij}=\sum_{n=1}^\infty \tilde{M}^{(n)}_{ij}$, which apparently is also transverse and traceless. The spatial metric in new coordinate system indeed takes the form of eqn. (\ref{Nonlinear SFSG}).
%%%%%%%%%%%%%%%%%%%%%%%%%%%%%%%%%%%%%%%%%%%%%
%%%%%%%%%%%%%%%%%%%%%%%%%%%%%%%%%%%%%%%%%%%%%
\section{Nonlinear Conserved $\zeta$ in $P(X)$ Fluids}\label{P(X) Fluid}
A perfect fluid free of vortex degrees of freedom has a dual low energy effective description involving only one scalar field $\psi$. In the cosmological context, the action is given by
\be \label{eff action PX}
S=S_{EH}+\int \sqrt{-g}\,P(X), \quad X\equiv -\pd_\mu\psi\pd_\nu \psi g^{\mu\nu}\,.
\ee
The energy momentum tensor can be obtained: 
\be
T_{\mu\nu}=2P'(X)\pd_\mu\psi\pd_\nu\psi+P(X)g_{\mu\nu}
\ee
Noticing that $u_\mu=\pd_\mu\psi X^{-\frac{1}{2}}$, $T_{\mu\nu}$ can be recast into the standard perfect fluid form (\ref{EM tensor Fluid}) if we identify 
\be \label{rho p u}
\rho=2 X P'-P\;, \qquad p=P\;.
\ee
Assuming that the background configuration of the scalar field is only time dependent $\langle\psi(x)\rangle=\psi_0(t)$ and that of the metric takes the usual FRW form, the Einstein equations and the scalar field equation of motion for unperturbed configurations read
\begin{align}
&3\mpl^2H^2=-P+2P'\dot{\psi}_0^2\label{Friedman 1}\\
&\mpl^2\dot{H}=-P' \dot{\psi}_0^2\label{Friedman 2}\\
&\left(P'+2P''\dot{\psi}_0^2\right)\ddot{\psi}_0+3HP'\ddot{\psi}_0=0\label{eom scalar PX}
\end{align}
Now let us fix the gauge. For the sake of defining a nonlinear $\zeta$, the unitary gauge is of most convenience: 
\be\label{UG PX}
\psi(x)=\psi_0(t)\;, \quad g_{ij}=a(t)^2e^{2\zeta}\exp\gamma_{ij}\;,
\ee

We shall show the $\zeta$ such defined is conserved outside the horizon. We adopt similar logic as before: first show that $\zeta$ =constant is a solution in the long wavelength limit, and then show that this solution is actually an attractor. 

Using the ADM variables, the effective action (\ref{eff action PX}) becomes 
\be\label{eff action PX ADM}
S=\int \N\sqrt{h}\left(\frac{\mpl^2}{2}R^{(3)}+\frac{\mpl^2}{2\N^2}(E^i_j\;E^j_i-E^2)+P(X)\right)
\ee
We can then obtain the constraint equations by varying this action with respect to $\N$ and $N^i$: 
\begin{align}
R^{(3)}&-\frac{1}{\N^2}\Bigg\{-6(H+\dot{\zeta})^2+4(H+\dot{\zeta})\hat{\na}_iN^i-\frac{1}{4}\Tr \left(\frac{d}{dt} e^{-\gamma}\frac{d}{dt}e^\gamma\right)+\frac{1}{2}\hat{\na}^iN_j\,\hat{\na}^jN_j\label{constraint 1 PX}\\
&+\frac{1}{2}\hat{\na}_jN^i\,\hat{\na}^jN_i-\hat{\na}_iN^i\,\hat{\na}_jN^j-\frac{1}{2}\left(e^{-\gamma}\frac{d}{dt}e^\gamma\right)^i_{\,j}\left(\hat{\na}^jN_i+\hat{\na}_iN^j\right)
\Bigg\}\nonumber\\
&+\frac{2}{\mpl^2}P\left(\frac{\dot{\psi}_0^2}{\N^2}\right)-\frac{4}{\mpl^2}\frac{\dot{\psi}_0^2}{\N^2}P'\left(\frac{\dot{\psi}_0^2}{\N^2}\right)=0\nonumber\\
\hat{\na}_i\Bigg[&-\frac{2}{\N}(H+\dot{\zeta})+\frac{1}{\N}\hat{\na}_jN^j\Bigg]+\hat{\na}_j\bigg[\frac{1}{2\N}\left(e^{-\gamma}\frac{d}{dt}e^\gamma\right)^j_i-\frac{1}{2\N}\hat{\na}_iN^j-\frac{1}{2\N}\hat{\na}^jN_i\bigg]=0\label{constraint 2 PX}
\end{align} 
From equation (\ref{constraint 2 PX}), we can see that $\hat{\na}_i N^j \sim \dot{\zeta}  \text{ (or } \dot{\gamma})$, i.e.~$\na_i N^j$ is of the zeroth order in the spatial gradient expansion ($\sigma=k/a H$). Thus solving $\delta \N\equiv \N-1$ via equation (\ref{constraint 1 PX}) up to order $\mathcal{O}(\sigma^0)$, we have
\be
\delta \N=-\mpl^2\left(P-P'\dot{\psi}_0^2+2P'' \dot{\psi}_0^4\right)^{-1}\left(3H\dot{\zeta}-H\hat{\na}_iN^i\right)+\mathcal{O}(\dot{\zeta}^2,\dot{\gamma}^2,\dot{\zeta}\dot{\gamma}).
\ee
Plugging this into the action (\ref{eff action PX ADM}) and expanding it to $\mathcal{O}(\sigma^0)$ while keeping all orders in fields,  we have 
\begin{align}
S&=-2\mpl^2\int \sqrt{h}\left\{3H^2+\dot{H}+\left(3H\dot{\zeta}-H\hat{\na}_iN^i\right)+\mathcal{O}(\dot{\zeta}^2,\dot{\gamma}^2,\dot{\zeta}\dot{\gamma})\right\}\nonumber\\
&=-2\mpl^2\int \frac{d}{d t}\left(a^3 e^{3\zeta}H\right)+\mathcal{O}(\dot{\zeta}^2,\dot{\gamma}^2,\dot{\zeta}\dot{\gamma})
\end{align}
where in the first equality we have used the background Einstein equations (\ref{Friedman 1}) (\ref{Friedman 2}) and the background scalar field equation (\ref{eom scalar PX}) and in the second equality we have neglected the contribution from $\hat{\na}_iN^i$ since it is a boundary term.  Therefore we conclude that the effective action in the long wavelength limit starts with terms involving two times derivatives (acting on different fields) and, by the same argument as before, that $\dot{\zeta}=0\,, \dot{\gamma}=0$ are solutions to the classical e.o.m.'s. 

Now we show that the solutions $\zeta=$ constant and $\gamma=$constant are attractors. As before we consider the fluctuations around these classical solutions and work out the quadratic action for $\delta \zeta$ and $\delta \gamma$.  Owing to the constancy of the classical solutions, the quadratic action for fluctuations take the same form as that for $\zeta$ and $\gamma$ themselves, which can be obtained by solving the constraint equations (\ref{constraint 1 PX}) (\ref{constraint 2 PX}) up to the linear order in fields, plugging into (\ref{eff action PX ADM}) the linearized solutions $\delta \N_1,\, N^i_1$ and expanding the effective action to the quadratic order. Thus we have
\be
\delta S\simeq \mpl^2\int a^3 \left[\frac{\varepsilon}{c_s^2}\dot{\delta \zeta}^2-\frac{\varepsilon}{a^2}(\pd_i\delta \zeta)^2\right]+\frac{1}{8}\left[\dot{\delta \gamma}_{ij}^2-\frac{1}{a^2}(\pd_i \delta\gamma_{jk})^2\right]
\ee
where $\varepsilon \equiv -\dot{H}/H^2$ and $c_s$ is the speed of sound defined as 
\be
c_s^2\equiv \frac{dp}{d\rho}\Big\vert _{X=\dot{\psi}_0^2}=\frac{P'(\dot{\psi}_0^2)}{P'(\dot{\psi}_0^2)+2\dot{\psi}_0^2P''(\dot{\psi}_0^2)}\;.
\ee
Thus in the long wavelength limit $\sigma\to 0$, the linearized equation for $\delta \zeta$ ($\delta \gamma$) possesses two general solutions, one being time independent and the other decaying, which confirms our claim that $\zeta=$ constant and $\gamma=$ constant are attractors. 

Before ending this section, let us remark that the nonlinear $\zeta$ constructed in $P(X)$ context agrees with that in $F(B)$ context. Notice that in unitary gauge (\ref{UG PX}),  $X=\dot{\psi}_0(t)^2(1+\delta \N)^{-2}$. Since on super-horizon scales $\delta \N \to 0$ due to the constancy of $\zeta$ and $\gamma$,  the constant time slices in the unitary gauge coincide with the uniform density slices (for $\rho$ is a function of $X$ only).  Similarly, since $N^i$ vanishes on large scales for the same reason,   we have $u^i\propto g^{0i}\propto N^i\to 0$, i.e.~the threading of the spatial coordinates is chosen such that the threads $x^i$ =constant coincide with the integral curves of the 4-velocity $u^\mu$ (the comoving worldlines).   Therefore $\zeta$ defined in $P(X)$ context is identical to $\zeta'$ in (\ref{new coo system}). 
%%%%%%%%%%%%%%%%%%%%%%%%%%%%%%%%%%%%%%%%%%%%%%%%%%%%%
%%%%%%%%%%%%%%%%%%%%%%%%%%%%%%%%%%%%%%%%%%%%%%%%%%%%%

\end{document}